\documentclass[aps,prl,reprint,showpacs,superscriptaddress,groupedaddress,amsmath,amssymb]{revtex4-1}
\usepackage{graphicx}
\usepackage{dcolumn}
\usepackage{bm}
\usepackage{color}
\usepackage{ulem}
\usepackage{epsfig}
\usepackage[english]{babel}

\begin{document}

\title{Spatio-temporal Control of Light Transmission\\ through a Multimode Fiber with Strong Mode Coupling}

\author{Wen Xiong$^{1}$, Philipp Ambichl$^{2}$, Yaron Bromberg$^{1}$, Brandon Redding$^{1}$, Stefan Rotter$^{2}$, Hui Cao$^{1}$}
\email{hui.cao@yale.edu}
\affiliation{$^{1}$Applied Physics Department, Yale University, New Haven CT 06520, USA \\
	$^{2}$Institute for Theoretical Physics, Vienna University of Technology, A-1040 Vienna, Austria}

\begin{abstract}
We experimentally generate and characterize the eigenstates of the Wigner-Smith time-delay matrix, called principal modes, in a multimode fiber with strong mode coupling. The unique spectral and temporal properties of principal modes enable a global control of the temporal dynamics of optical pulses transmitted through the fiber, despite random mode mixing. Our analysis reveals that the well-defined delay time of the eigenstates are formed by multi-path interference, which can be effectively manipulated by the spatial degrees of freedom of the input wavefront. This study is essential to controlling the dynamics of wave scattering, paving the way for coherent control of pulse propagation through complex media. 
\end{abstract}

\maketitle

The temporal dynamics of wave scattering in complex systems has been widely studied in quantum mechanics, nuclear physics, acoustics and optics. Most of these studies, e.g., electromagnetic or ultrasonic wave propagation in billiards \cite{Doron90,Savin03,Fyodorov05,Stefan11}, electron transport through quantum dots \cite{Brouwer97,Brouwer99}, and light scattering in random media \cite{Lagendijk96, Azi99, Tiggelen99,Shi15}  focused on the statistics of delay times, i.e.,  eigenvalues of the Wigner-Smith time-delay matrix \cite{Eisenbud48,Wigner55,Smith60}. Despite innumerable  trajectories  the wave could take through an open complex system, an eigenstate of the Wigner-Smith matrix remarkably has a well-defined delay time. Some of the eigenstates are particlelike with their wavefunctions concentrating on a single trajectory \cite{Stefan11}, and hence a definite transit time is expected. Most of the states, however, consist of enormous trajectories with various lengths such that it is concealed how well-defined delay times can be attributed to these states.

Largely in parallel, the Wigner-Smith eigenstates were introduced for multimode optical fibers (MMFs), which attracted much attention in recent years due to the rapid development of space-division multiplexing for telecommunications \cite{RichardsonNatPho13}. Inherent imperfections and external perturbations introduce random coupling of the guided modes in the fiber and cause temporal broadening and distortion of transmitted pulses. As a generalization of principal states of polarization in a single-mode fiber \cite{Poole86}, the Wigner-Smith eigenstates, also called principal modes (PMs) of MMFs, were proposed to suppress modal dispersion \cite{Fan05}. 

Advances in wavefront shaping techniques now make it possible to probe a single Wigner-Smith eigenstate in optics. Recently PMs were observed experimentally  in a few-mode fiber with weak mode coupling \cite{Carpenter15}. In this regime, mode coupling in the fiber is only perturbative and hence the PMs are similar to the eigenmodes of a perfect fiber. In the strong mode coupling regime, however, all modes are strongly mixed and the multiple scattering of light between different guided modes generates numerous paths for light to propagate through the fiber. It remains obscure how  PMs are formed with well-defined delay times and what properties they possess in the presence of non-perturbative mode mixing. Here we report on a demonstration of PMs in a MMF with strong mode coupling. Our analysis uncovers that the well-defined delay time of a PM can be explained by multi-path interference that is tailored by spatial degrees of freedom of the input wavefront. This multi-path interference also determines the spectral bandwidth of PMs, which limits the temporal width of optical pulses that can be transmitted  through the fiber without distortion.

\begin{figure}
\includegraphics[scale=0.58]{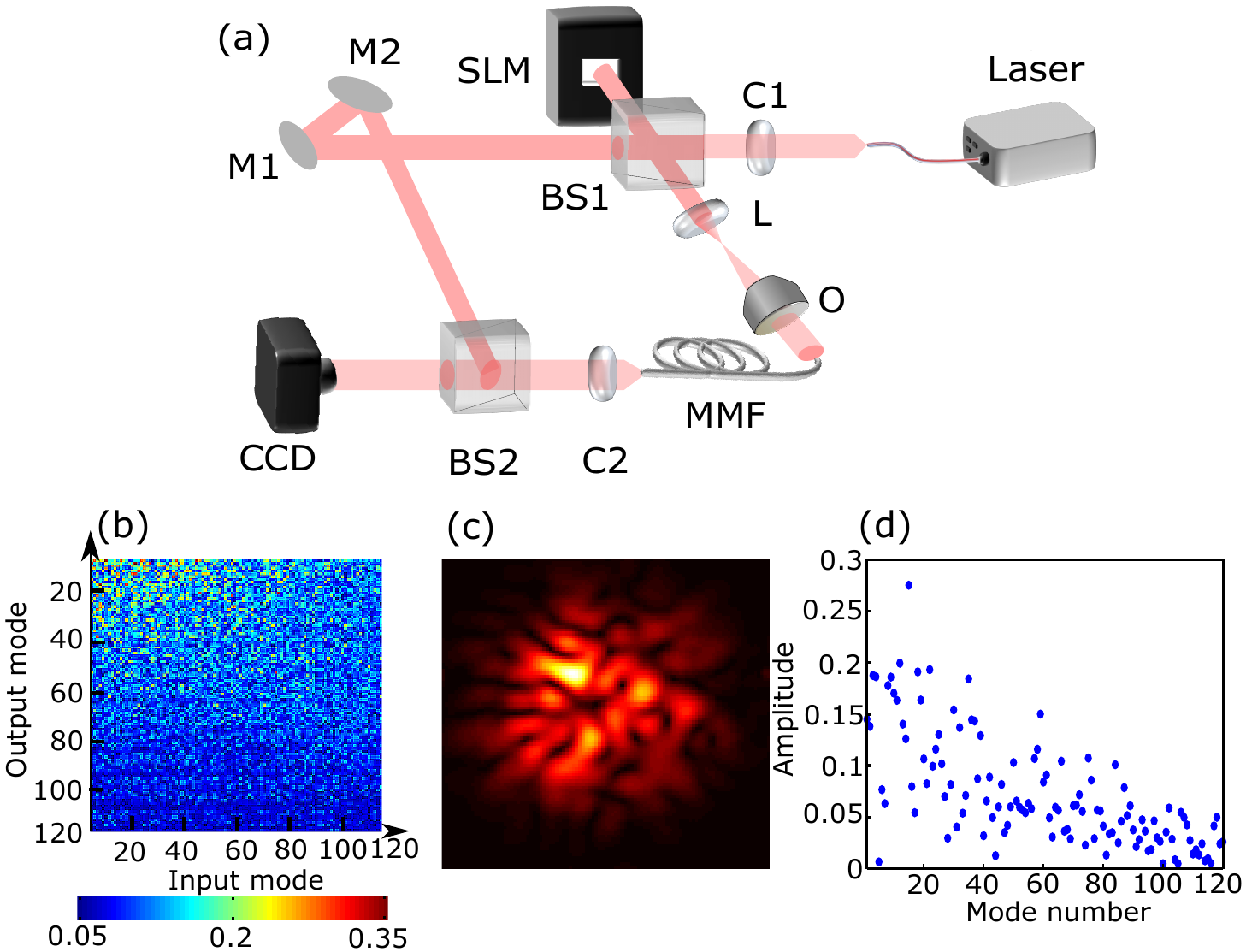}\caption{(color online) 
	(a) Schematic of the simplified experimental setup for measurement of the transmission matrix of a MMF. The continuous-wave output from a tunable laser source is collimated by a collimator (C1), and split into two arms by a beam splitter (BS1). The light in one arm is modulated by the SLM and imaged to the fiber facet by a lens (L) and an objective (O). The output field from the fiber is collimated (C2) and combined with the light in the other arm at a second beam-splitter (BS2). By offsetting the BS2 to introduce a phase tilt between the two wavefronts, interference fringes are formed. From the interferogram recorded by the camera (CCD), the output field is extracted. The mirrors (M1, M2) are used to match the path-length of the two arms of the interferometer. The MMF is one meter long with 50 $\mu$m core diameter and 0.22  numerical aperture. 
	(b) Amplitude of the measured transmission matrix  at $\lambda$ = 1550 nm ($\omega$=1219 THz). 
	(c) Amplitude profile of the output field of a PM.
	(d) Decomposition of the PM in (c) by the linearly polarized modes, revealing that it consists of many modes.  
\label{fig:Principal-mode}}
\end{figure}

The Wigner-Smith time-delay matrix is defined as $Q \equiv -iS^{-1} {dS}/{d\omega}$, where $S$ is the scattering matrix of a system \cite{Wigner55,Smith60}. In the absence of backscattering in the fiber, it can be expressed as $Q \equiv -iT^{-1} {dT}/{d\omega}$ \cite{Fan05, Juarez12}, in which $S$ is replaced by the transmission matrix $T$. We experimentally measure the field transmission matrix of a MMF as a function of frequency in an off-axis holographic setup shown schematically in Fig.~\ref{fig:Principal-mode}(a). To introduce strong mode coupling in a one-meter-long fiber, we apply stress to the fiber using clamps. The field transmission matrix is measured in momentum space and converted to the mode basis. Figure~\ref{fig:Principal-mode}(b) shows the amplitudes of a measured transmission matrix. Whichever mode the input light is launched into, the output field spreads over all modes, although higher order modes have lower amplitude due to stronger loss. The transmission matrix is very different from that in the weak coupling regime, which has larger elements closer to the matrix diagonal, confirming the modes are strongly coupled in the current fiber. 

\begin{figure}
\includegraphics[scale=0.4]{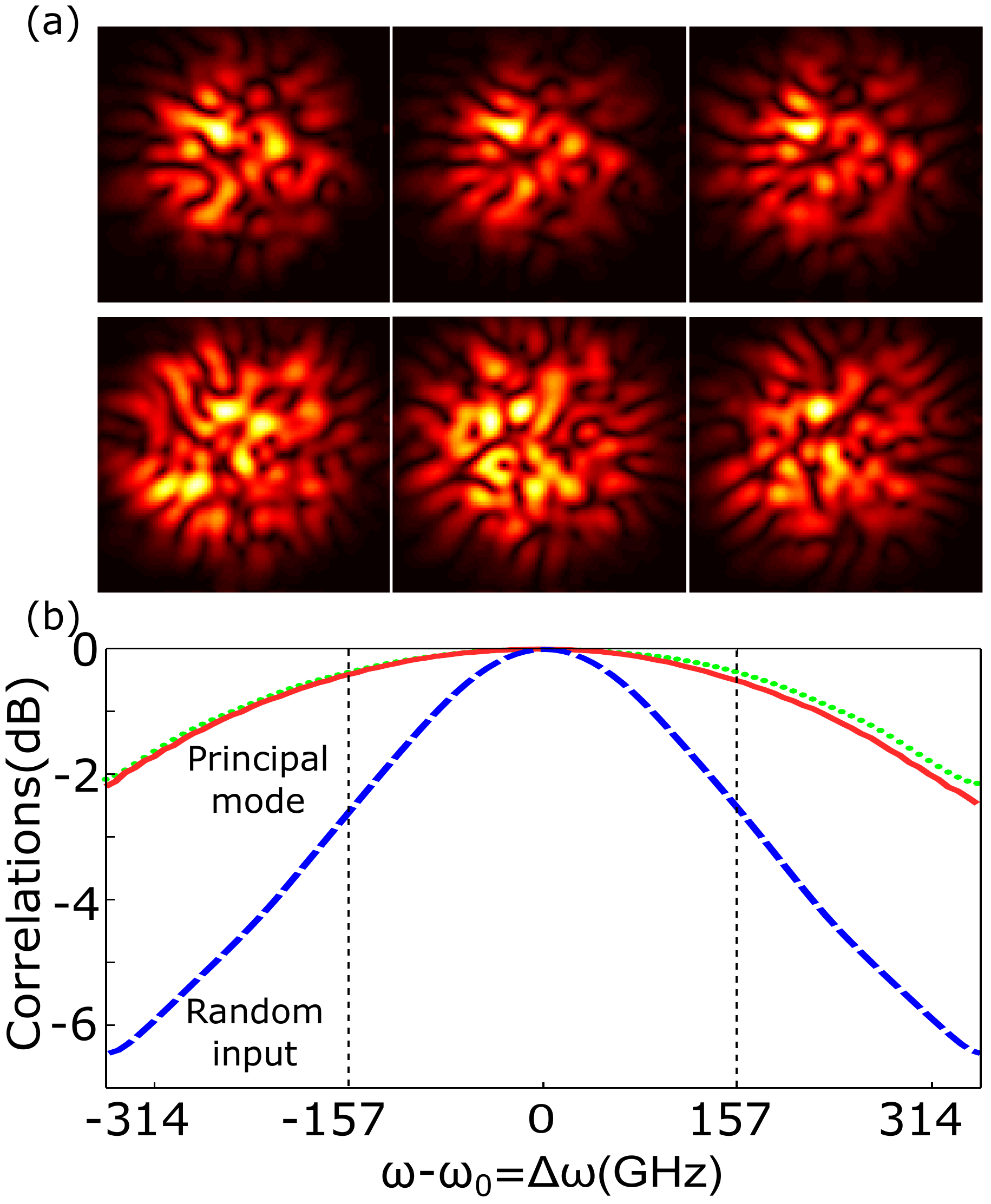}\caption{(color online)  
	 (a) Output field amplitude for the input wavefront of a PM at $\omega_0$ = 1219 THz (top row), or a random superposition of linearly polarized modes (bottom row). The input frequency is $\omega - \omega_0$ =  $-157$ GHz (left column),  $0$ (middle column), and $157$ GHz (right column). The output field patterns for the PM input are similar while those for random input are totally different. 
	 (b) Spectral correlation function $C(\Delta \omega)$ of the output field pattern, measured experimentally for a PM (red solid curve), or calculated from the measured transmission matrix and input spatial profile of the same PM (green dotted curve). For comparison, $C(\Delta \omega)$ for a random input is also shown (blue dashed curve). $C(\Delta \omega)$ is normalized to one at $\Delta \omega=0$. The agreement between the red and green curves illustrates the accuracy of the measurement. The output field pattern for the PM decorrelates much slower with frequency than the random input, and $C(\Delta \omega)$ displays a plateau at $\Delta \omega=0$. 
	 \label{fig:Spectral-Characteristics}}
\end{figure}

The eigenstates of $Q$, which is constructed from transmission matrices measured at different frequencies, give the input fields of  PMs. They are unique input states that produce frequency-independent output field patterns to the first order of frequency variation \cite{Fan05}. In an ideal MMF, the PMs are simply the linearly polarized modes, which are the eigenmodes of the fiber in the weak guiding approximation. In the weak mode coupling limit, the fiber length is less than the correlation length (the distance beyond which the spatial field profile becomes uncorrelated \cite{HoJLT14}), and each PM consists of a few modes with similar propagation constants. However, if the fiber length well exceeds the correlation length, all modes are thoroughly mixed and the PMs are expected to be distinct from those in the weak coupling regime. We use a spatial light modulator (SLM) to generate input wavefronts of  individual PMs of this MMF with strong mode coupling. To modulate both amplitude and phase of the input field with a phase-only SLM, a computer-generated phase hologram is employed \cite{Arrizon07}. Figure~\ref{fig:Principal-mode}(c) shows the output pattern of a PM, which is speckled and does not resemble any linearly polarized mode of the fiber. Modal decomposition of the pattern reveals that the PM is a mixture of many modes [Fig.~\ref{fig:Principal-mode}(d)], in contrast to the PM in the weak mode coupling regime. 

To investigate the spectral property of PMs, we scan the frequency $\omega$ while keeping the input field pattern to that of a PM at $\omega_0$. The output field pattern is measured at each frequency and compared to that at $\omega_0$. Figure~\ref{fig:Spectral-Characteristics}(a) shows the far-field patterns at three frequencies (top row), and they are nearly identical. For comparison, a random superposition of modes at the input results in  different output profiles at these three frequencies [bottom row of Fig.~\ref{fig:Spectral-Characteristics}(a)]. This striking difference illustrates that the output field pattern of the PM decorrelates much slower with frequency. 

To be more quantitative, we calculate the spectral correlation function $C(\Delta \omega \equiv \omega - \omega_0) \equiv |{\bf \Psi}(\omega_0)^* \cdot {\bf \Psi}(\omega)|$, where ${\bf \Psi}(\omega)$ is a vector representing the output fields in all spatial channels with its magnitude normalized to unity. As shown in Fig.~\ref{fig:Spectral-Characteristics}(b), $C(\Delta \omega)$ for the PM is significantly larger than that for the random input. It displays a broad plateau at $\Delta\omega=0$. To understand the shape of the correlation curve, we denote $C(\Delta\omega) = \mbox{cos}[\theta(\Delta\omega)]$, where $\theta$ is the angle between the two output field vectors at  $\omega$ and $\omega_0$. Since $\theta(0) = 0$, the first-order derivative of $C$ with respect to $\Delta \omega$ vanishes at $\Delta \omega=0$ for any input wavefront. The second-order derivative at $\Delta \omega=0$ is proportional to $[\theta'(0)]^{2}$, where $\theta' \equiv d \theta / d \Delta \omega$. For the PM, $\theta'(0) = 0$, because the output spatial profile remains unchanged to the first order of frequency variation. Thus the second-order derivative vanishes for the PM, leading to a plateau of the correlation curve, that is absent for the random input.

\begin{figure}
\includegraphics[scale=0.335]{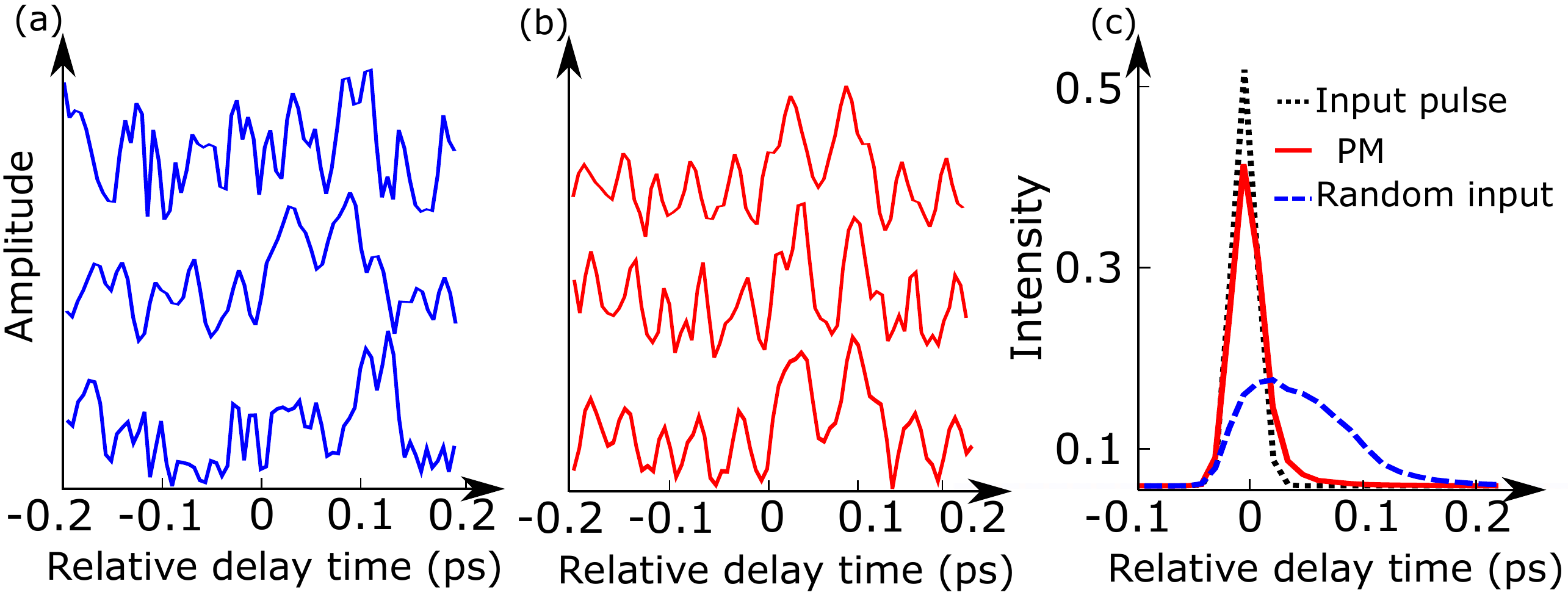}
\caption{(color online)   
	(a,b) Temporal variation of the output field amplitude in three spatial channels (three speckles grains) when an optical pulse is launched into a random superposition of fiber modes (a) or a PM at $\omega_0$ = 194 THz (b). The spatial profile of the output field is recorded in a frequency range of 400 GHz with a step size of 2.5 GHz. The Fourier transform is then performed to obtain the field evolution in time. The temporal traces of individual spatial channels are totally different for the random input, but nearly identical for the PM input. (c) Spatially integrated intensity of the input (black dotted curve) and the output pulses when a Gaussian pulse is injected to the MMF with random spatial profile (blue dashed curve) or with the profile of a PM (red solid curve). \label{fig:temporal}}
\end{figure}

In a next step we probe the temporal dynamics of a single PM. Like other open chaotic systems, the transmission of a pulse through a MMF with strong mode coupling involves spatial and temporal  distortions. Strong mode mixing results in hopping of the light among modes with different propagation constants. Thus the light can take many paths of varying lengths through the fiber. The output in each spatial channel (e.g. speckle grain) is a sum of waves with different paths, each associated with a respective time delay, leading to temporal broadening and distortion of the input pulse. Typically, the temporal trace varies from one channel to another, since the combination of paths differs. This is confirmed by simulating the propagation of a pulse, $\phi(t)=\int \phi(\omega) \, e^{-i\omega t} d\omega$, with a field spectrum $\phi(\omega)$. The pulse is launched into a random superposition of modes at the input of the fiber, and the output field patterns are recorded experimentally as a function of frequency. We perform the Fourier transform to obtain the temporal evolution of the output field in each spatial channel. Figure~\ref{fig:temporal}(a) shows the temporal traces of field magnitude in three spatial channels. They are very different from each other, due to strong mode scrambling in the fiber. 

However, if the input light is coupled to a PM, the output fields in different spatial channels are synchronized, as shown in Fig.~\ref{fig:temporal}(b). This is a direct consequence of the invariance of output field pattern with frequency. Namely, the output field vector at frequency $\omega$ can be written as ${\bf \Psi}(\omega) = \phi(\omega) T(\omega) {\bf \Phi}$, where the input field vector ${\bf \Phi}$ corresponds to a PM at $\omega_0$. If the input bandwidth is less than the spectral correlation width of the PM, $T(\omega) {\bf \Phi} \approx \alpha(\omega) \hat{\bf \Psi}_0$, where $\hat{\bf \Psi}_0$ is a unit vector representing the normalized output field profile for the PM at $\omega_0$, and $\alpha(\omega)$ is a complex number that may vary with frequency. The Fourier transform of ${\bf \Psi}(\omega)$ gives the output field vector ${\bf \Psi}(t) = \tilde{\phi}(t) \hat{\bf \Psi}_0$, where  $\tilde{\phi}(t) = \int \phi(\omega) \alpha(\omega) e^{-i \omega t} dt$ represents the output pulse shape. Hence, the spatial and temporal variations of the output field become decoupled for the PM. The temporal traces in all output channels are identical up to a constant factor given by the elements of $\hat{\bf \Psi}_0$. The spatial profile of the output field remains constant in time, allowing the spatial and temporal distortions to be corrected separately. For example, the output pulse shape can be tailored by modulating the spectral phase of input spectrum $\phi(\omega)$. Since the output fields are spatially coherent, a spatial mask can convert the output to any desired pattern or focus to a diffraction-limited spot. 

Let us consider a simple case, $\alpha(\omega) \simeq \alpha_0 e^{i \eta(\omega)}$, where $\alpha_0$ is a constant amplitude and the phase $\eta(\omega) \simeq \eta(\omega_0) + \eta'_0 \, (\omega-\omega_{0})$, where $\eta'_0$ is the value of ${d\eta}/{d\omega}$ at $\omega_0$. Then the output pulse, $\tilde{\phi}(t) \propto \phi(t-\eta'_0)$, has the same temporal shape as the input one. This is confirmed by synthesizing a pulse with Gaussian spectrum and flat phase at the input. The output intensity, summed over all spatial channels, is plotted in Fig.~\ref{fig:temporal}(c) together with the input pulse intensity. The output pulse has negligible broadening and shape distortion, despite strong mode coupling in the fiber. In contrast, the same pulse, but with a random input pattern, suffers from severe broadening as seen in Fig.~\ref{fig:temporal}(c). PMs can thus compensate for the temporal distortion induced by modal dispersion in an MMF.   

\begin{figure}
\includegraphics[scale=0.5]{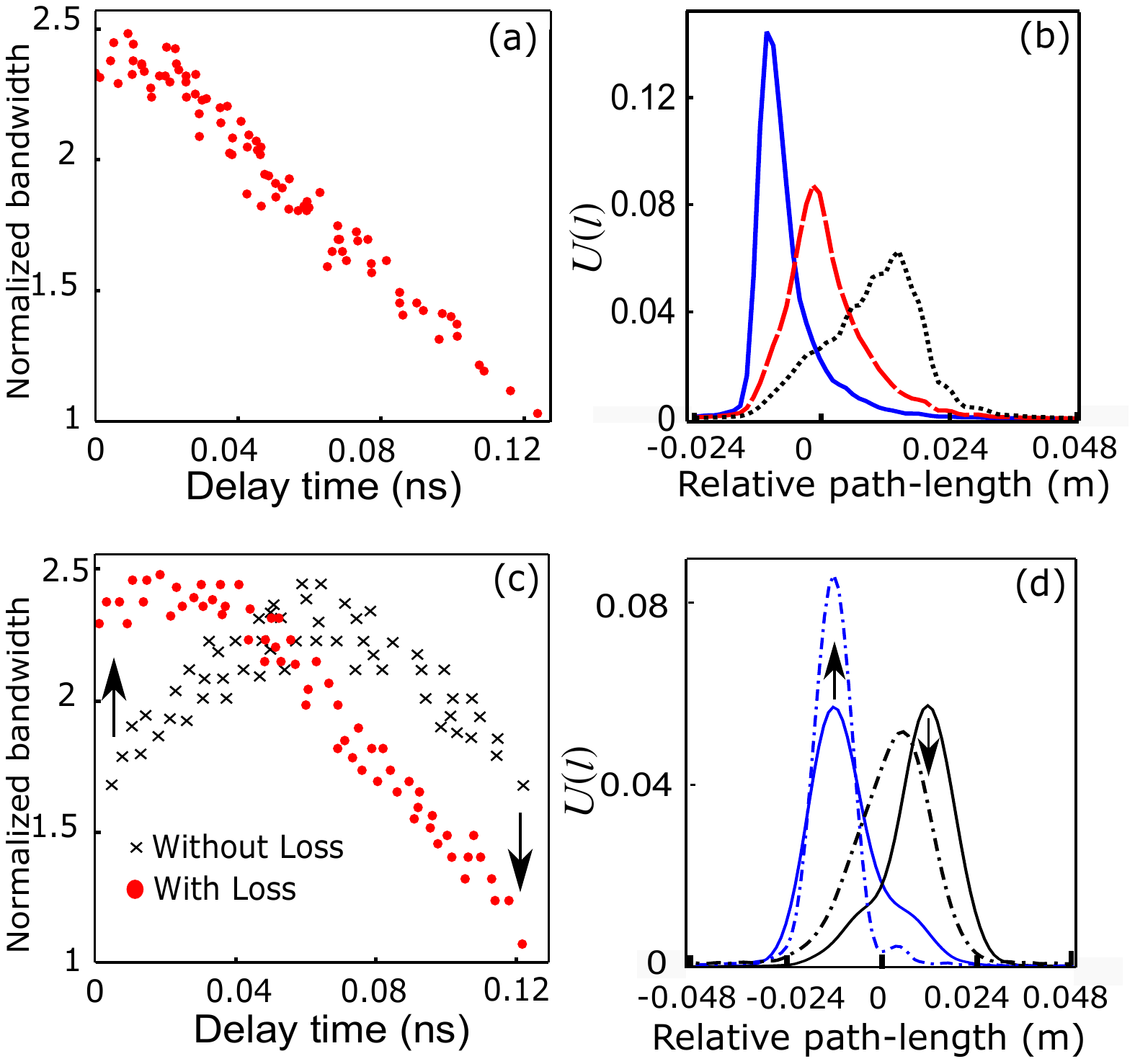}
\caption{(color online) 
	(a) Measured spectral correlation width  $\Delta \omega_c$ of  PMs with different delay times. $\Delta \omega_c$, given by $|C(\Delta\omega_c)| = 0.9|C(0)|$, is normalized by the average bandwidth of random inputs. The shortest delay time is set to 0.  	
	(b) Intensity distribution over the path-length $U(l)$ for three measured PMs with the delay time = 0, 0.06, 0.12 ns.  The relative path-length $l$ is obtained by subtracting the average path-length of random inputs. $U(l)$ is normalized: $\int U(l) \, dl = 1$. 	
	(c) Calculated spectral correlation width of the PMs with (red circles) and without (black crosses) mode-dependent loss.  	
	(d) Intensity distribution over the path-length for two  PMs with the delay time = 0 ns (blue line), 0.12 ns (black line), in the presence (dashed line) or absence (solid line) of mode-dependent loss. The mode-dependent loss narrows (broadens) the path-length distribution for the fast (slow) principle mode, thereby increasing (reducing) the spectral correlation width.  
\label{fig:bandwidth}}
\end{figure}

The unique spectral and temporal properties of PMs hold only within a finite frequency range. It is hence important to determine the bandwidth of PMs.   Since the spectral decorrelation of the output pattern for any input wavefront depends on the fiber properties, such as the fiber length and numerical aperture, we consider below the ratio of the bandwidth of the PM to the average bandwidth of random inputs. Figure \ref{fig:bandwidth}(a) plots the experimentally measured bandwidth of all PMs versus their delay time. The shorter the delay, the larger the bandwidth. 

To obtain a physical understanding of  PMs and their bandwidth, we resort to the intuitive picture of optical paths in the fiber. The output field pattern is a result of interference of waves following innumerable possible trajectories in the MMF created by strong mode coupling. As the input frequency changes, the relative phase accumulated along trajectories of different length varies, modifying the output field pattern. More specifically, the output field in the $m$-th spatial channel can be written as $\Psi_m = \int u_m(l) d l$, where $u_m(l)$ is a sum of fields taking all possible paths with the same length $l$. With a small frequency detuning $\Delta \omega$, the output field becomes $\Psi_{m}(\Delta\omega) = \int u_m(l)e^{i\Delta\omega \, l/c}dl$, in the weak guiding approximation. Thus $u_m(l)$ can be obtained experimentally from the Fourier transform of $\Psi_{m}(\Delta\omega)$. Accounting for all spatial channels, $U(l) = \sum |u_m(l)|^2$ gives the intensity distribution over the path-length, which is determined by the input wavefront. How fast the output field decorrelates is determined by how broad the intensity is distributed over the path-length spectrum. The narrower the distribution, the weaker the dephasing among different path-lengths by frequency detuning, and the smaller the change in the interference pattern at the output. Hence, the PMs with shorter delay times have larger spectral correlation widths due to the narrower path-length distribution.   

Figure \ref{fig:bandwidth}(b) compares $U(l)$ for three PMs with different delay times. The fast PM has intensity concentrated on shorter paths. Although the waves can take many longer paths, the destructive interference of different trajectories with the same length makes $U(l)$ vanish for the longer path-length. The opposite happens to the slow PM. The redistribution of intensity among different path-length is determined by the input wavefront. Therefore, the delay time in a MMF with strong mode coupling is determined by the multi-path interference effect, which can be effectively controlled by spatial degrees of freedom of the input wavefront. 

The final question we address here is why the fast PM has a narrower path-length distribution. To answer this question, we perform numerical simulations using the concatenated waveguide model \cite{Ho11}. For simplicity, we consider a planar waveguide with a 300$\mu$m core and a 0.22 numerical aperture, supporting 86 guided modes. The one-meter-long waveguide is composed of 20 segments, in each of which light propagates without mode coupling. Between adjacent segments, the guided modes are randomly coupled, as simulated by a unitary random matrix. To include mode-dependent loss, we introduce a uniform absorption coefficient to each waveguide segment. The higher-order modes, with smaller propagation constants, have longer transit time, thus experiencing more attenuation. In terms of optical path, the longer paths have more loss than the shorter ones. 

Figure~\ref{fig:bandwidth}(c) plots the bandwidth of all PMs in the absence of mode-dependent loss (black crosses). The fast and slow PMs have almost identical bandwidth, as the corresponding intensity distribution among the path-length exhibits similar spread at different mean values [Fig.~\ref{fig:bandwidth}(d), solid curves]. With the introduction of mode-dependent loss, the bandwidth of fast PMs increases, while the bandwidth of slow ones decreases, leading to agreement with the experimental data [compare red dots in Fig.~\ref{fig:bandwidth}(a) with those in Fig.~\ref{fig:bandwidth}(c)]. This rearrangement can be explained by the change in intensity distribution over the path-length, $U(l)$, that is plotted in Fig.~\ref{fig:bandwidth}(d). The distribution for a fast PM, which concentrates on short paths, becomes narrower, because the longer paths are further suppressed by the loss. For the slow PM, the stronger attenuation of longer paths not only shifts the peak of $U(l)$ to smaller $l$, but also broadens the distribution. Qualitatively, the change of PM bandwidth induced by the mode-dependent loss is not sensitive to the kind of loss the fiber experiences, as long as  higher-order modes have more loss, as expected for the MMF.

In summary, we experimentally probe individual eigenstates of the Wigner-Smith time-delay matrix of a multimode fiber with strong mode coupling. We find that the well-defined delay times of the eigenstates are formed by multi-path interference, which can be manipulated by the spatial degrees of freedom of the input wavefront. The multi-path interference also determines the frequency range over which the unique spectral and temporal properties of the Wigner-Smith eigenstates preserve.  Within the bandwidth, the spatial and temporal variations of the transmitted field are decoupled for the eigenstates, enabling a global spatio-temporal control of pulse transmission through complex media. Such global control, which is more challenging than the control over a single spatial channel such as spatio-temporal focusing \cite{Katz11,McCabe11,Aulbach11,Aulbach12,Shi13,Morales15}, has potential applications to optical communication, imaging and sensing.

We acknowledge Joel Carpenter and Nicolas Fontaine for interesting discussions. 
This work is supported partly by the US National Science Foundation under the Grant Nos. ECCS-1509361 and DMR-1205307. P.A. and S.R. acknowledge support by the Austrian Science Fund (FWF) through projects SFB NextLite (F49-P10) and project GePartWave(I1142).

\end{document}